\documentclass[preprint,showpacs,preprintnumbers,amsmath,amssymb]{revtex4}
\usepackage{epsfig}
\usepackage{graphicx}
\usepackage{dcolumn}
\usepackage{bm}

\newcommand{\ord}{{\cal O}}
\def\beq{\begin{equation}}
\def\eeq{\end{equation}}
\def\eeqn{\end{equation}}
\newcommand\iden{\leavevmode\hbox{\small1\normalsize\kern-.33em1}}

\newcommand{\bea} {\begin{eqnarray}}
\newcommand{\eea} {\end{eqnarray}}

\let\jnfont=\rm
\def\NPB#1,{{\jnfont Nucl.\ Phys.\ B }{\bf #1},}
\def\PLB#1,{{\jnfont Phys.\ Lett.\ B }{\bf #1},}
\def\EPJC#1,{{\jnfont Eur.\ Phys.\ Jour.\ C }{\bf #1},}
\def\PRD#1,{{\jnfont Phys.\ Rev.\ D }{\bf #1},}
\def\PRL#1,{{\jnfont Phys.\ Rev.\ Lett.\ }{\bf #1},}
\def\MPLA#1,{{\jnfont Mod.\ Phys.\ Lett.\ A }{\bf #1},}
\def\JPG#1,{{\jnfont J.\ Phys.\ G }{\bf #1},}
\def\CTP#1,{{\jnfont Commun.\ Theor.\ Phys.\ }{\bf #1},}
\def\JHEP#1,{{\jnfont JHEP \ }{\bf #1},}
\def\NPPS#1,{{\jnfont Nucl.\ Phys.\ Proc.\ Suppl.\ }{\bf #1},}
\def\CPC#1,{{\jnfont Computl.\ Phys.\ Commun.\ }{\bf #1},}
\def\CPL#1,{{\jnfont Chin.\ Phys.\ Lett. }{\bf #1},}
\def\APPB#1,{{\jnfont Acta\ Phys.\ Polon.\ B }{\bf #1},}

\def\lsim{\raise0.3ex\hbox{$<$\kern-0.75em\raise-1.1ex\hbox{$\sim$}}}
\def\gsim{\raise0.3ex\hbox{$>$\kern-0.75em\raise-1.1ex\hbox{$\sim$}}}

\begin{document}

\title{\ \\[10mm]
Dark matter in little Higgs model
under current experimental constraints from LHC, Planck and Xenon}

\author{Lei Wang$^1$, Jin Min Yang$^{2}$, Jingya Zhu$^2$}

\affiliation{
$^1$ Department of Physics, Yantai University, Yantai 264005, PR China\\
$^2$ State Key Laboratory of Theoretical Physics,\\
     Institute of Theoretical Physics, Academia Sinica,
             Beijing 100190, PR China
\vspace{0.5cm} }


\begin{abstract}
We examine the status of dark matter (heavy photon) in the littlest
Higgs model with T-parity (LHT) in light of the new results from the
LHC Higgs search, the Planck dark matter relic density and the
XENON100 limit on the dark matter scattering off the nucleon. We
obtain the following observations: (i) For the LHC Higgs data, the
LHT can well be consistent with the CMS results but disfavored by
the ATLAS observation of diphoton enhancement; (ii) For the dark
matter relic density, the heavy photon in the LHT can account for
the Planck data for the small mass splitting of mirror lepton and
heavy photon; (iii) For the dark matter scattering off the nucleon,
the heavy photon can give a spin-independent cross section below the
XENON100 upper limit for $m_{A_H}>95$ GeV ($f> 665$ GeV); (iv) A fit
using the CMS Higgs data gives the lowest chi-square of 2.63 (the SM
value is 4.75) at
 $f\simeq$ 1120 GeV and $m_{A_H}\simeq$ 170 GeV (at this point
the dark matter constraints from Planck and XENON100 can also be satisfied).
Such a best point and its nearby favored region (even for a $f$ value
up to 3.8 TeV) can be covered by the future XENON1T (2017) experiment.
\end{abstract}

\pacs{12.60.Fr, 95.35.+d, 14.80.Ec}

\maketitle

\section{Introduction}
To solve the fine-tuning problem of the standard model (SM), the
little Higgs theory \cite{LH} is proposed as a kind of electroweak
symmetry breaking mechanism accomplished by a naturally light Higgs
boson. The littlest Higgs model (LH) \cite{lst} provides an
economical realization for this theory.
Further, to relax the constraints from the electroweak precision data
\cite{cstrnotparity}, a discrete symmetry called T-parity is
introduced to the LH \cite{tparity,lhti}. The LH with T-parity (LHT)
predicts a heavy photon as a candidate for the weakly interacting
massive particle (WIMP) dark matter (DM), whose relic density,
direct detection, indirect detection and phenomenology at the LHC
have been intensively studied \cite{lhdm1,lhdm}.

Very recently, some experiments have made significant progress,
which allow for a test for new physics like the LHT.
On the one hand, for the dark matter the Planck collaboration \cite{planck}
released its relic density as $\Omega_c h^2\pm\sigma=0.1199\pm0.0027$ (in
combination with the WMAP data \cite{wmap}) and the CDMS II
direct detection experiment has reported three WIMP-candidate events
corresponding to a WIMP around 8.6 GeV \cite{cdms}.
However, such a CDMS result is in tension with other direct
detection results like the latest XENON100 results \cite{xenon},
which provided the most stringent upper limits on the spin-independent
WIMP-nucleon scattering cross section for a WIMP above 7 GeV.

On the other hand, for the Higgs search
the CMS and ATLAS collaborations have announced
observation of a Higgs-like boson around 125 GeV
\cite{cmsh,atlh,1303-a-com, 1303-c-com}. This observation is
supported by the Tevatron search which showed a 3.1$\sigma$
excess at $M_h=125$ GeV \cite{1303-t-com}. The properties of this
observed particle are well consistent with the SM Higgs boson
for most of the search channels. Note that the Higgs diphoton rate from the
ATLAS is sizably larger than the SM expectation, $1.6\pm 0.3$
\cite{1303-a-com}, but the central value of CMS is smaller than the SM
prediction, $0.77\pm0.27$ \cite{1303-c-com}. The Higgs properties in the
LHT have been studied in \cite{lhhrr,lhtiyuan,13010090,yue} and
the diphoton rate was found to be always suppressed (in contrast to
the low energy supersymmetric models which can either enhance or
suppress the diphoton rate \cite{cao}).

In this work we examine the status of dark matter (heavy photon) in
the LHT under the latest experimental constraints from the LHC Higgs
result, the Planck DM relic density and the XENON100 (2012) limit on
the DM-nucleon scattering. In Sec. II we recapitulate the dark
matter sector in the LHT. In Sec. III we examine the status of dark
matter (heavy photon) in light of the  latest experimental results.
Finally, we give our conclusion in Sec. IV.

\section{The littlest Higgs model with T-parity}

This model \cite{lst} consists of a nonlinear sigma model with a
global $SU(5)$ symmetry which is broken down to $SO(5)$ by a vacuum
expectation value (VEV) $f$. A subgroup $[SU(2) \otimes U(1)]^2$ of
$SU(5)$ is gauged. T-parity is an automorphism which exchanges the
$[SU(2) \otimes U(1)]_1$ and $[SU(2) \otimes U(1)]_2$ gauge fields.
While all the SM particles are T-even,
the new gauge bosons ($W_H^\pm$, $Z_H$, $A_H$) and the triplet scalar
($\Phi^{++}$, $\Phi^+$, $\Phi^0$, $\Phi^P$) are T-odd, whose masses
are given by
\beq
m_{Z_H} \simeq m_{W_H} = gf(1-\frac{v^2}{8f^2}),
  \qquad
m_{A_H} \simeq \frac{g^{\prime}f}{\sqrt{5}}(1-\frac{5v^2}{8f^2}),\qquad
        m_{\Phi}\simeq \sqrt{2}m_h\frac{f}{v}.
\label{ahvv2}
\eeq
Here $h$ and $v$ are respectively the SM-like Higgs boson and
its vacuum expectation value (vev). The relation between $G_F$ and
$v$ is modified from its SM form and reads as \cite{lhtiyuan}
\beq
v\simeq v_{SM}(1+\frac{1}{12}\frac{v^2_{SM}}{f^2}),
\eeq
where
$v_{SM}=246$ GeV is the SM Higgs vev. The heavy photon $A_H$ is
typically the lightest T-odd state and thus can serve as a candidate
for dark matter.

In the top quark sector, there are a T-even (denoted as $T$) and a
T-odd partner (denoted as $T_-$). The T-even one mixes with
the top quark and cancels the quadratic divergent contribution of
the top quark to the Higgs boson mass. The mixing can be parameterized
by
\beq
r=\frac{\lambda_1}{\lambda_2},~~
c_t=\frac{1}{\sqrt{r^2+1}},~~ s_t=\frac{r}{\sqrt{1+r^2}},
\eeq
where
$\lambda_1$ and $\lambda_2$ are two dimensionless  top
quark Yukawa couplings.
The masses of the T-even partner  and the
T-odd partner  are given by
\bea
m_T&=&\frac{m_tf}{s_t c_t
v}\left[1+\frac{v^2}{f^2}(\frac{1}{3}-s_t^2 c_t^2)\right],\nonumber \\
m_{T_-}&=&\frac{m_tf}{s_t
v}\left[1+\frac{v^2}{f^2}(\frac{1}{3}-\frac{1}{2}s_t^2
c_t^2)\right].
\eea

For each SM quark (lepton), a heavy mirror quark (lepton) with T-odd
quantum number is added in order to preserve T-parity. Their masses
are given by
\bea
&&m_{u_{Hi}}=\sqrt{2}\kappa_{qi}f(1-\frac{v^2}{8f^2}),\qquad
m_{d_{Hi}}=\sqrt{2}\kappa_{qi}f,\nonumber\\
&&m_{v_{Hi}}=\sqrt{2}\kappa_{li}f(1-\frac{v^2}{8f^2}),\qquad
m_{l_{Hi}}=\sqrt{2}\kappa_{li}f,
\eea
where $\kappa_{qi}$ and $\kappa_{li}$ with $i=1, 2, 3$
are the eigenvalues of the mirror quark
and lepton mass matrices, respectively.

For the SM down-type quarks (leptons), the Higgs couplings
of LHT have two different cases \cite{lhtiyuan}:
\begin{eqnarray}
\frac{C_{hd\bar{d}}}{C_{hd\bar{d}}^{\rm SM}}
&\simeq&1-\frac{1}{4}\frac{v_{SM}^2}{f^2}+\frac{7}{32}
\frac{v_{SM}^4}{f^4} ~~~~{\rm for~LHT-A}, \label{Higgs-downA} \nonumber\\
&\simeq&1-\frac{5}{4}\frac{v_{SM}^2}{f^2}-\frac{17}{32}
  \frac{v_{SM}^4}{f^4} ~~~~{\rm for~LHT-B}.\nonumber
\label{eq15}
\end{eqnarray}
The relation of down-type quark couplings also applies to the lepton
couplings.

In our analysis we use MicrOMEGAs3.2 to calculate the relic density
and the cross section between DM and nucleon \cite{micromegas}. The
CalcHEP LHT model files are provided by \cite{0609179}. We add the
Higgs couplings to the u-quark, d-quark and electron, and modify the
$Z$ and $W$ couplings to mirror fermions. In addition, we assume the
interactions between the mirror fermions and the SM fermions are
diagonal.

Some typical Higgs and DM couplings are given by
\cite{lhtiyuan,0610298,mirrorfr} \bea
\begin{array}{ll}
hA_H A_H:~~-\frac{g'^2}{2}v\left[1-\frac{v^2}{f^2}
(\frac{4}{3}-\frac{2c_W}{s_W}x_H)\right]g^{\mu\nu},~~~~&
hu_{Hi}\bar{u}_{Hi}:~~\frac{m_{u_{Hi}}}{v}\frac{v^2}{4f^2}, \\
hW^+W^-:~~\frac{2m_W^2}{v}(1-\frac{1}{6}\frac{v^2}{f^2})g^{\mu\nu},&
hZZ:~~\frac{2m_Z^2}{v}(1-\frac{1}{6}\frac{v^2}{f^2})g^{\mu\nu},\\
hW_H^+ W_H^-:~~-\frac{2m_{W_H}^2}{v}\frac{v^2}{4f^2}g^{\mu\nu},&
  h\Phi^+\Phi^-:~~\frac{2m_{\Phi}^2}{v}\frac{v^2}{3f^2}, \\
hu\bar{u}:~~-\frac{m_u}{v}(1-\frac{2}{3}\frac{v^2}{f^2}),&
hc\bar{c}:~~-\frac{m_c}{v}(1-\frac{2}{3}\frac{v^2}{f^2}),\\
ht\bar{t}:~~-\frac{m_t}{v}\left[1+\frac{v^2}{f^2}(-\frac{2}{3}+c_t^2
s_t^2)\right],&
hT\bar{T}:\frac{m_T}{v}c_t^2 s_t^2\frac{v^2}{f^2},\\
A_Hl_{i}\bar{l}_{Hi}:~~-(\frac{g'}{10}-\frac{g}{2}x_H\frac{v^2}{f^2})\gamma^{\mu}P_L
,&
A_H\nu_{i}\bar{\nu}_{Hi}:-(\frac{g'}{10}
+\frac{g}{2}x_H\frac{v^2}{f^2})\gamma^{\mu}P_L\\
A_Hd_{i}\bar{d}_{Hi}:~-(\frac{g'}{10}-\frac{g}{2}x_H\frac{v^2}{f^2})\gamma^{\mu}P_L
,&
A_Hu_{i}\bar{u}_{Hi}:~~-(\frac{g'}{10}+\frac{g}{2}x_H
\frac{v^2}{f^2})\gamma^{\mu}P_L\\
A_Ht\bar{u}_{H3}:~~-\left[\frac{g'}{10}+(\frac{g'}{20}s_t^2-\frac{g}{2}x_H)
\frac{v^2}{f^2}\right]\gamma^{\mu}P_L,&
Wl_{Hi}\bar{\nu}_{Hi}:~~\frac{g}{\sqrt{2}}\gamma^{\mu}
\left[1-(\frac{v^2}{8f^2}P_R)\right]\\
Z\nu_{Hi}\bar{\nu}_{Hi}:~~\frac{g}{c_W}\gamma^{\mu}
\left[\frac{1}{2}-(\frac{v^2}{8f^2}P_R)\right],&
Zl_{Hi}\bar{l}_{Hi}:\frac{g}{c_W}\gamma^{\mu} (-\frac{1}{2}+s_W^2),
\end{array}
\label{coupling}
\eea
where $x_H=\frac{5gg'}{4(5g^2-g'^2)}$.

\section{Dark matter in LHT under current experimental constraints}
\subsection{Implication of LHC Higgs data on LHT parameter space}
In our calculations, the Higgs mass is fixed as 125.5 GeV, and the
new free parameters are $f,~r,~\kappa_{li},~\kappa_{qi}$. The
electroweak precision data favor $f >$ 500 GeV and 0.5 $<r<$ 2
\cite{500}. We assume that the three generations of mirror quarks (leptons)
are degenerate in mass, namely $\kappa_{l1}=\kappa_{l2}=\kappa_{l3}$
and $\kappa_{q_1}=\kappa_{q_2}=\kappa_{q_3}$. For
$\kappa_{qi}<0.45$, the mirror quark is lighter than the heavy gauge
bosons $W_H$ and $Z_H$, and thus its only decay  mode is the two-body
decay into $A_H$ and a SM quark.
 The ATLAS and CMS collaborations have
analyzed jets plus missing transverse momentum signal, and not yet
found any hints of new physics \cite{atlasjet,cmsjet}. Therefore,
we take $0.45< \kappa_{qi}< 1$ conservatively. In addition, we impose
the LEP limits on the masses of charged leptons which are required
to be larger than 105 GeV \cite{105}.

We consider the relevant QCD and electroweak corrections using the
code Hdecay \cite{hdecay}. For the Higgs productions and decays, the
LHT gives the corrections by directly modifying the Higgs couplings
to the relevant SM particles. For the loop-induced decays $h \to gg$
and $h\to\gamma\gamma$, the LHT gives the partial corrections via
the reduced $ht\bar{t}$ and $hWW$ couplings, respectively. Besides,
$h\to gg$ can get contributions from the loops of heavy partner
quark $T$ and mirror up-type quarks. In addition to the loops of the
heavy quarks involved in the $h\to gg$, the decay $h\to
\gamma\gamma$ can be also altered by the loops of $W_H$, $\Phi^\pm$
and $\Phi^{\pm\pm}$ in the LHT. The doubly charged $\Phi^{\pm\pm}$
contributions are enhanced by a relative factor 4 in the amplitude,
but can still be ignored due to the very small coupling
$h\Phi^{++}\Phi^{--}$ (in contrast to the type II seesaw model whose
doubly charged scalar can give the dominant contributions to the
decay $h\to\gamma\gamma$ \cite{htm}). Since the mirror charged
lepton, the mirror down-type quark and the top parter $T_-$ do not
have tree-level couplings to the Higgs boson, they do not contribute
to $h \to gg$ and $h\to\gamma\gamma$ at leading order. For
$m_h=125.5$ GeV, the decay $h\to A_H A_H$ is kinematically forbidden
in the LHT (such an invisible decay was possible in some
supersymmetric models \cite{cao-yang-dm}).

The decays $h \to gg$ and $h\to\gamma\gamma$ are not sensitive to the
 mirror quark masses as long as they are much
larger than half of the Higgs boson mass. The parameter $r$
determines the Higgs couplings to $t$, $T$ and $m_T$, and is
involved in the calculations of $h\to gg$ and $h\to
\gamma\gamma$. The $r$ dependence of the top quark loop and $T$
quark loop can cancel to a large extent, as can be seen from Eq.
(\ref{coupling}). Therefore, the Higgs signal rates in many channels
are only sensitive to the scale $f$.

\begin{figure}[tb]
 \epsfig{file=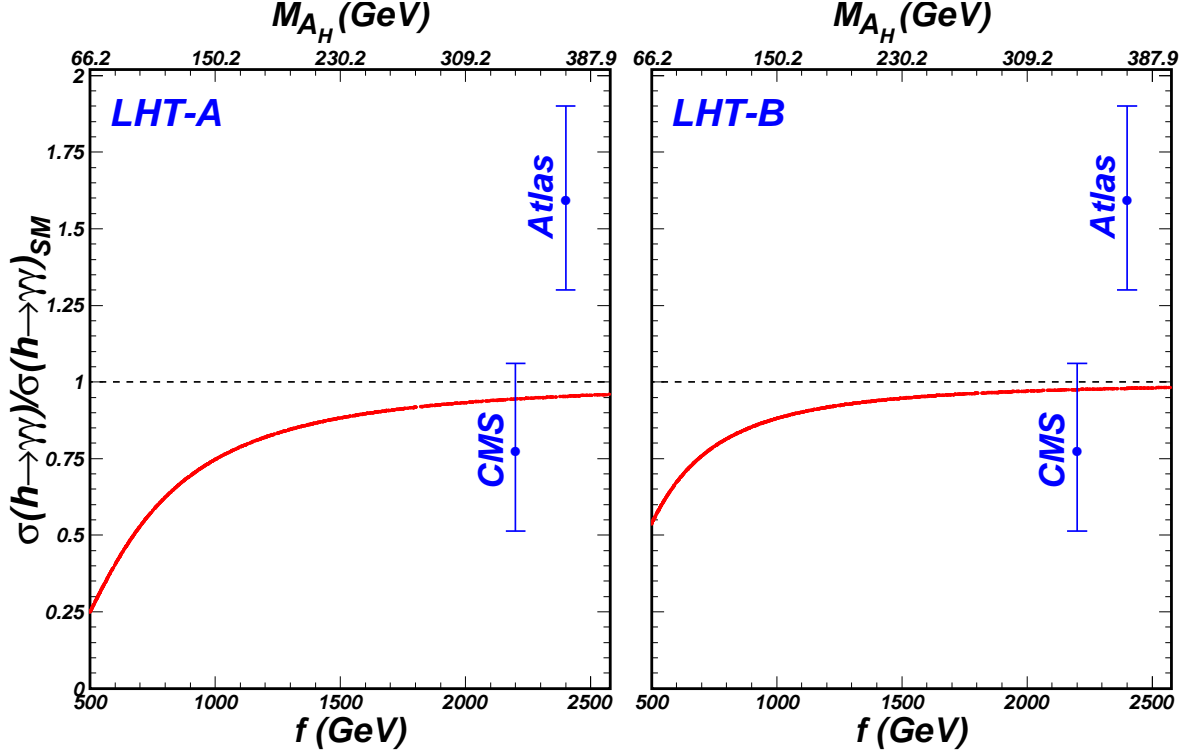,height=10cm}
\vspace{-0.4cm} \caption{The scatter plots of the LHT parameter space
projected on the plane of the LHC diphoton rate versus $f$.
The inclusive diphoton data are taken from \cite{1303-a-com,1303-c-com}.}
\label{fighrr}
\end{figure}

Requiring that the heavy photon is the lightest T-odd particle, we
scan over the parameter space of $f$, $r$, $\kappa_{qi}$ and
$\kappa_{li}$ in the ranges allowed by the electroweak precision data
($\kappa_{li}$ is not involved in the calculation of the Higgs signal rates).
We show the inclusive diphoton signal rate normalized to the SM
value in Fig. \ref{fighrr}. From this figure we find that the
diphoton rates in the LHT-A and LHT-B are always suppressed, and
approach to the SM predictions for a large scale $f$. The
suppression in the former is more sizable than in the latter because
the $hb\bar{b}$ coupling in the LHT-B is suppressed more sizably.
Since the ATLAS diphoton data is above the SM value by about
$2\sigma$, the predicted rates in both the LHT-A and LHT-B are
outside the $2\sigma$ range of the ATLAS data. For the CMS diphoton
data which shows no enhancement relative to the SM value, the LHT-A
and LHT-B can both give the signal rates in its $1\sigma$ range. In
the following, we will focus on the CMS data instead of combining
the two groups' results.

\begin{figure}[tb]
 \epsfig{file=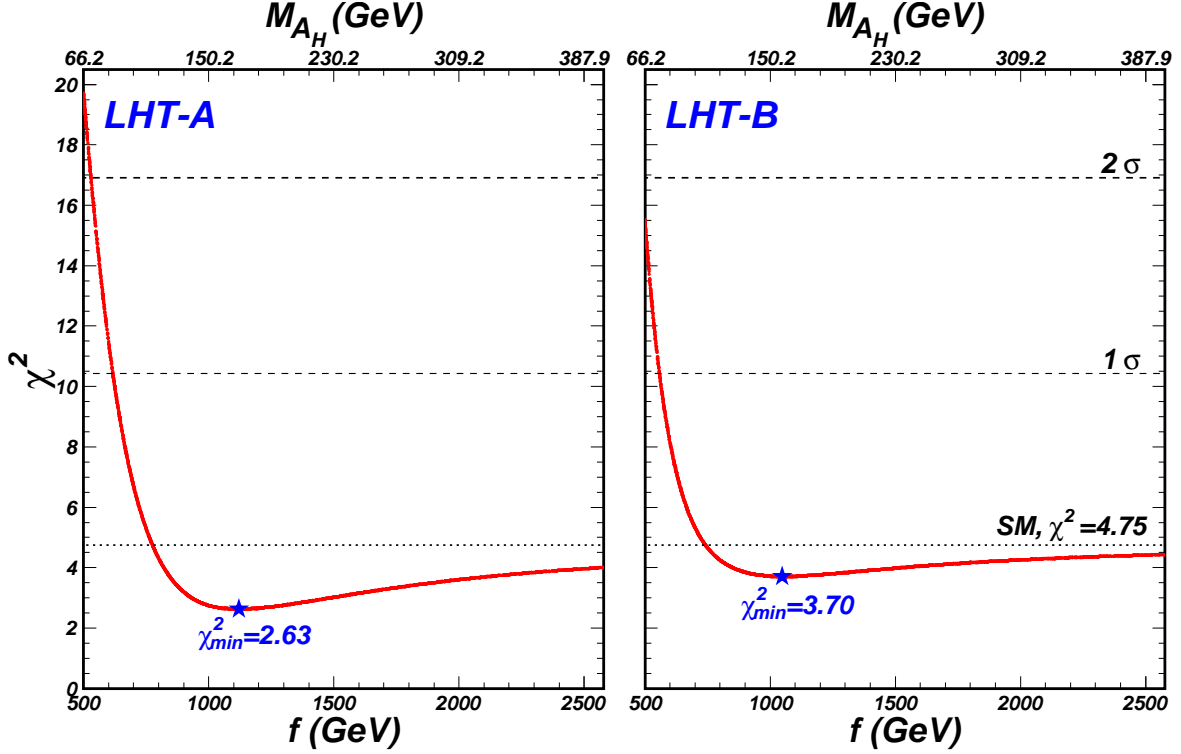,height=10cm}
\vspace{-0.4cm}
\caption{The scatter plots of the LHT parameter space
projected on the plane of $\chi^2$ versus $f$.
Here we considered the CMS Higgs data in 9 channels.
The samples with the minimal values of $\chi^2$ are marked
out as stars. Also showed are
the $1\sigma$ ($\chi^2=10.43$) and $2\sigma$ ($\chi^2=16.92$
) values as well as the SM fit value ($\chi^2=4.75$).}
\label{chi}
\end{figure}

Now we perform a fit to the CMS Higgs data in the
LHT-A and LHT-B. We compute the $\chi^2$ values by the method
introduced in \cite{method,hfit-ita}, with the CMS
Higgs data in 9 channels from Fig.4 of \cite{1303-c-com}.
Since the data for
different exclusive search channels presented by one collaboration
are not independent, we consider the correlation coefficient
as in \cite{1212-Gunion, 1307-Gunion}.

In Fig. \ref{chi} we project the samples on the plane of $\chi^2$
versus $f$. Similar to the diphoton rates, the $\chi^2$ of LHT-A and
LHT-B is only sensitive to $f$. The $\chi^2$ is larger than the SM
value for low values of $f$, then becomes smaller than the SM value
for intermediate values of $f$, and finally approaches to the SM
value for sufficiently high values of $f$. For $f$ around 1 TeV,
$\chi^2$ in the LHT-A and LHT-B reaches to the minimal value, which
is $2.63$ for LHT-A and $3.70$ for LHT-B. So we see that the best
point favored by the CMS Higgs data is at $f\sim 1$ TeV.

We also performed the fit using the ATLAS data
\cite{1303-a-com,1303-cos}. We found that the $\chi^2$ values are
much larger than using the CMS data. The main source of large
$\chi^2$ comes from the diphoton enhancement \cite{1303-a-com}.

\subsection{Dark matter relic density and scattering with nucleon}
Our results show that the heavy photon relic density and its
spin-independent cross section with nucleon are very similar for the
LHT-A and LHT-B. So we only present the results for LHT-A. We will
display the dark matter relic density and its spin-independent cross
section with the nucleon in the parameter space allowed by the CMS
Higgs data at $2\sigma$ level (as shown in the left panel of Fig. 2,
most samples in our scan can survive such a $2\sigma$ criterion).
The theoretical predictions in the LHT-A will be compared with the
relic density data from the Planck and the scattering rate limit
from the XENON100. Also, the future XENON-1T sensitivity will be
shown for the LHT-A.

The heavy photon pair-annihilation processes include $A_HA_H\to
f\bar{f}$, $ZZ$, $WW$ which proceed via an $s$-channel $h$ exchange,
and $A_HA_H\to hh$ which proceeds via a 4-point contact interaction,
an $s$-channel $h$ exchange, and $t$- and $u$-channel $A_H$
exchange. Also, $A_HA_H\to f\bar{f}$ can proceed via the $t$- and
$u$-channel T-odd fermion exchange (including the mirror quark,
mirror lepton, top partner $T_-$), whose contributions to the relic
density are generally suppressed by the interactions between the
T-odd fermions and SM fermions mediated by the heavy photon. In
addition, the mirror lepton can have an important effect on the
relic density via the coannihilation processes for the mirror lepton
masses close to the heavy photon. However, the other T-odd
particles, including the mirror quarks, top partner quark $T_-$,
heavy gauge bosons and scalars, do not contribute to the relic
density since their mass are much larger than $A_H$.

\begin{figure}[tb]
 \epsfig{file=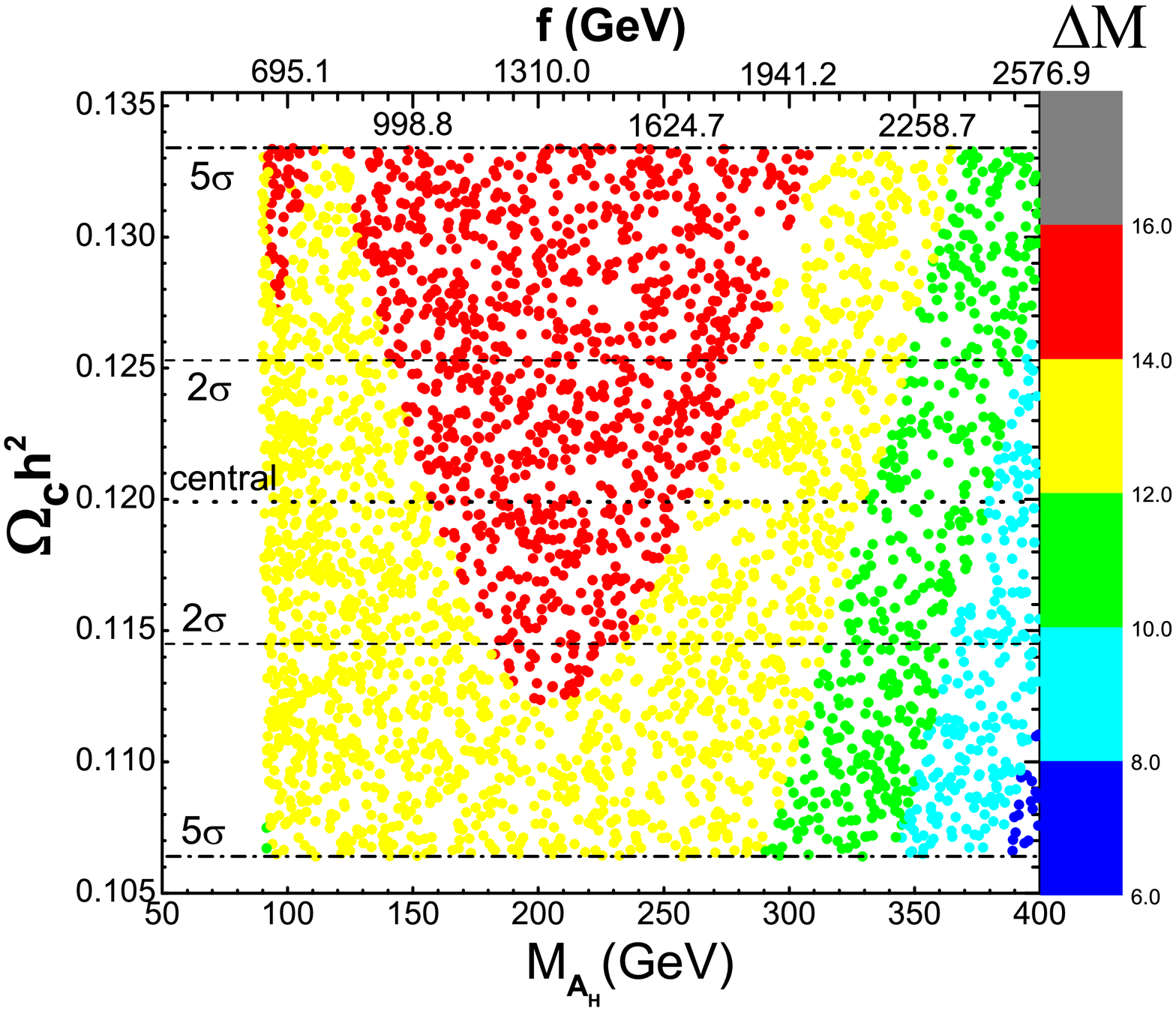,height=7.0cm}
  \epsfig{file=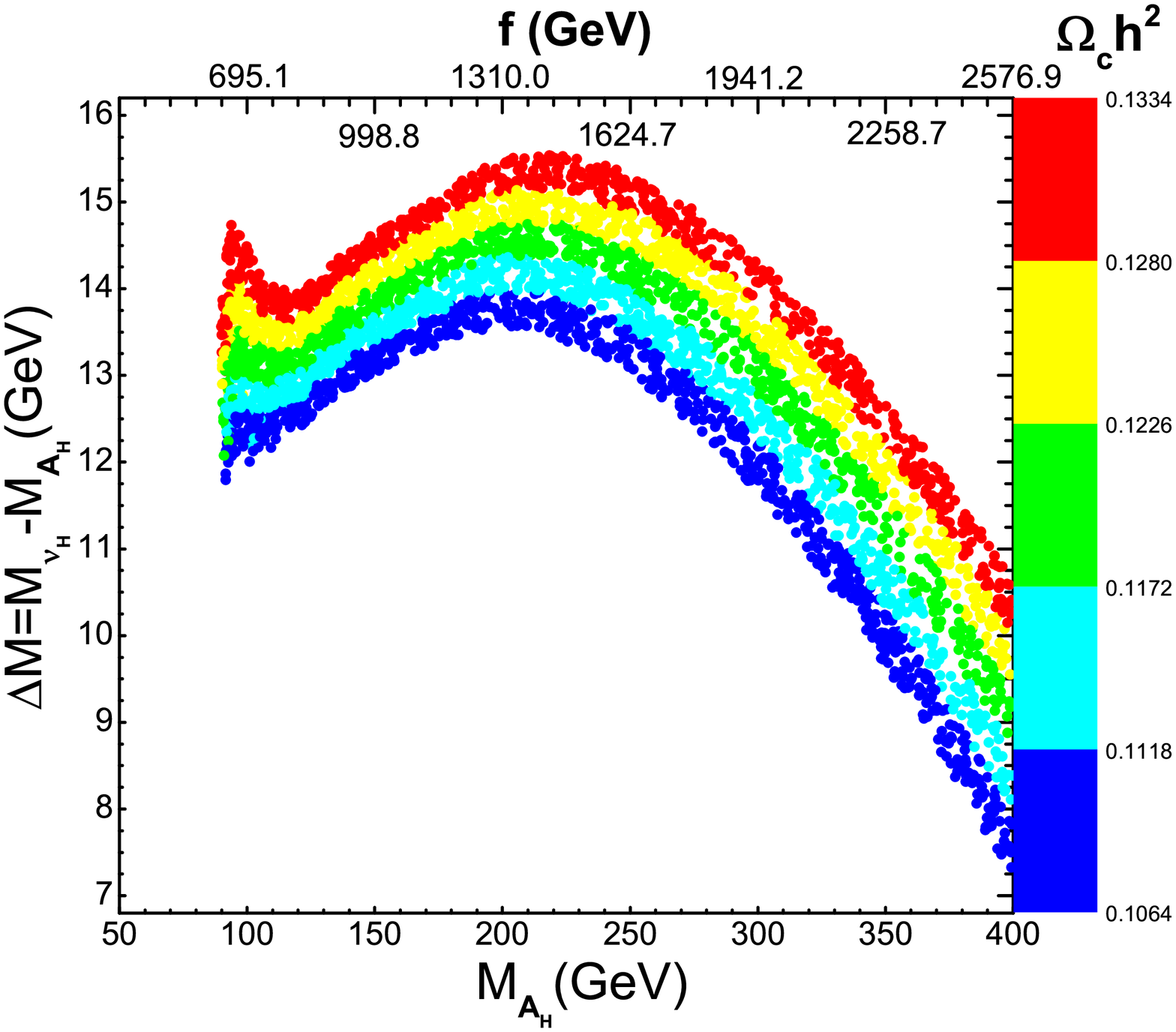,height=7.0cm}
\vspace{-0.4cm} \caption{The scatter plots of the LHT-A parameter
space allowed by the CMS Higgs data at $2\sigma$ level, showing the
dark matter relic density $\Omega_ch^2$. The central value of 0.1199
is from the Planck data \cite{planck}.} \label{omega}
\end{figure}

In Fig. \ref{omega} we project the LHT samples  showing the
dependence of the heavy photon relic density on $m_{A_H}$ and
$\Delta M$ (the mass splitting between mirror neutrino and heavy
photon). We see that in order to account for the DM relic density,
$\Delta M$ must be small and thus the mirror leptons play an
important role via the coannihilation processes. For the heavy
photon pair-annihilation, there is no $s$-channel Higgs resonance
since the mass splitting of $2m_{A_H}$ and $m_h$ is much larger than
the total width of Higgs. Further, the relevant Higgs and heavy
photon couplings are suppressed by a factor of
$1-\ord{(\frac{v^2}{f^2})}$ (see Eq. 6). Therefore, the cross
sections of the heavy photon pair-annihilation are too small to
provide the correct relic density of DM, and the mirror leptons have
to play an important role via the coannihilation processes.

\begin{figure}[tb]
 \epsfig{file=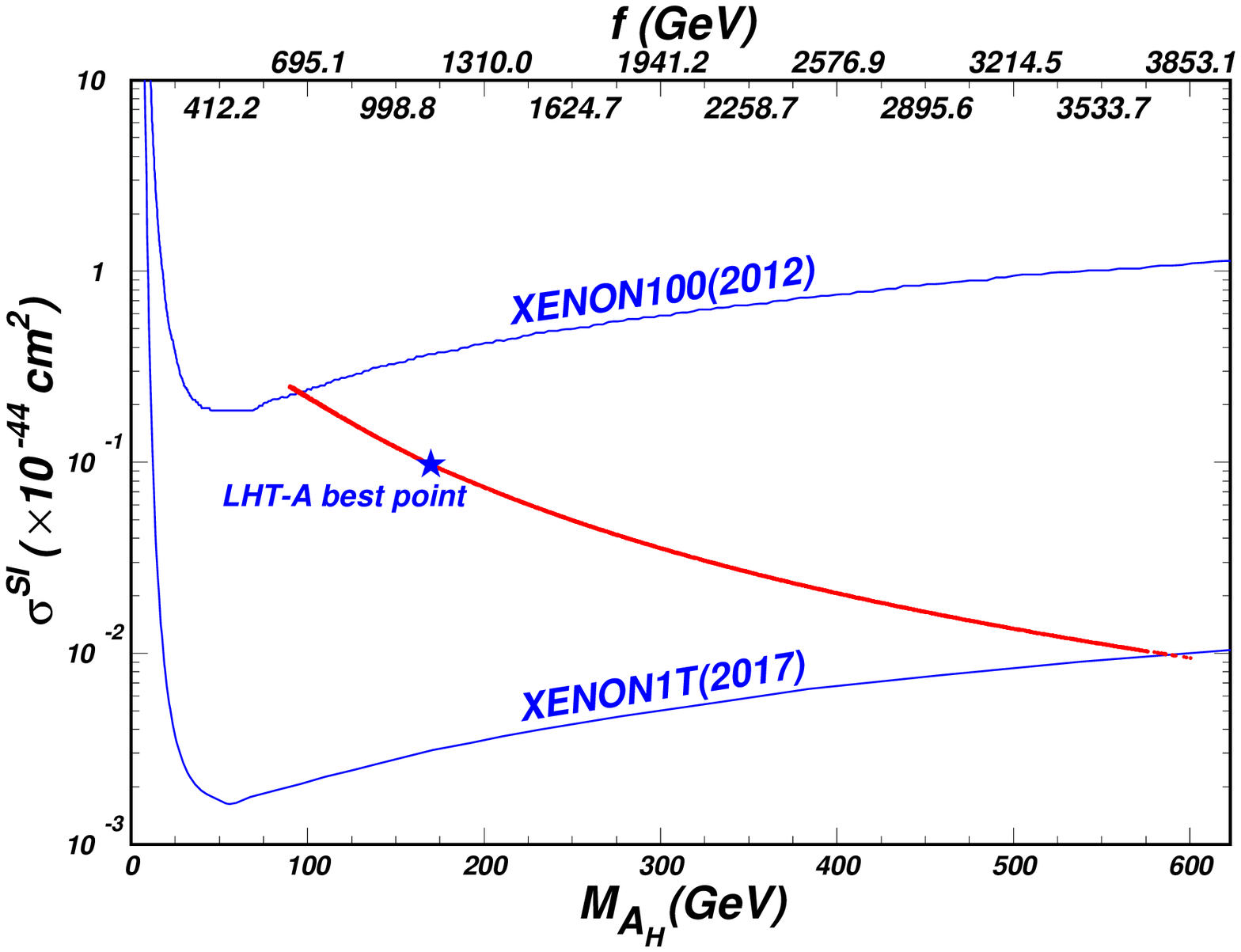,height=10cm}
\vspace{-0.4cm} \caption{ The scatter plots of the LHT-A parameter
space allowed by the CMS Higgs data at $2\sigma$ level and by the
Planck dark matter relic density  at $2\sigma$ level, showing the
spin-independent scattering cross section off the nucleon. The best
point, which gives minimal $\chi^2$ for the CMS Higgs data and also
the best relic density (closest to the measured central value), is
marked as a star. The curves denote the XENON100 (2012) limits
\cite{xenon} and XENON1T (2017) sensitivity \cite{XENON1T},
respectively.} \label{xenon}
\end{figure}

The heavy photon scattering off the nucleon can occur via
exchanging a Higgs boson or a mirror quark.
The former will give the dominant contribution to the spin-independent
cross section, especially for the Higgs-gluon interaction via the
heavy quark loops. For the latter case, since the mirror quarks are
much heavier than the heavy photon, we have no enhancement
in the propagator \cite{lhdm1}. They can not contribute
sizably to the spin-independent cross section due to the
small couplings of $A_H u \bar{u}_{H1}$ and $A_H d \bar{d}_{H1}$.

In Fig. \ref{xenon} we display the scatter plots of the LHT-A
parameter space allowed by the CMS Higgs data at $2\sigma$ level and
by the Planck dark matter relic density  at $2\sigma$ level, showing
the spin-independent scattering cross section off the nucleon. We
see that the spin-independent cross section decreases with the
increasing of $m_{A_H}$, and is below the upper limit of XENON100
(2012) for $m_{A_H}>95$ GeV ($f > 665$ GeV).
 The best point from the fit of
the CMS Higgs data (also give the relic density in the $2\sigma$
range) happens at $f\simeq$ 1120 GeV and $m_{A_H}\simeq$ 170 GeV.
Such a best point and its nearby favored region (even for a $f$
value up to 3.8 TeV) can be covered by the future XENON1T (2017)
experiment.

Finally, in Table \ref{best} we present the detailed information for
two samples (LHT-A P1 and LHT-B P3) which give minimal $\chi^2$ for
the CMS Higgs data and also the best relic density (closest to the
measured central value). Also, another two samples (LHT-A P2 and
LHT-B P4) for $f=$ 800 GeV are given for comparison. As previously
discussed, $\chi^2$ is sensitive to $f$ and the relic density is
sensitive to $f$ and $\kappa_l$. For the best-fit points,
$\sigma^{SI}$ is almost the same in the LHT-A and LHT-B, while the
Higgs properties have sizable differences. Especially, the rates for
the Higgs signals $\gamma\gamma$, $ZZ^*$, and $WW^*$ via the
$VBF+VH$ production channel are enhanced in the LHT-B, but
suppressed in the LHT-A, which may be useful for distinguishing
between the two models.

\begin{table}
\vspace{-1.5cm} \caption{The detailed information of some samples in
the LHT-A and LHT-B.}
\renewcommand{\arraystretch}{0.96}
  \setlength{\tabcolsep}{2pt}
  \centering
  \begin{tabular}{|c|c|c|c|c|}
    \hline \hline
     &~~  LHT-A P1 ~~& ~~LHT-A P2~~ &~~ LHT-B P3~~&~~ LHT-B P4~~\\
    \hline
     $f (GeV)$
     & 1121.5 & 800.0 &  1050.7  &  800.0
     \\
     r & 1.908 & 1.183 &  0.504  &  1.183
     \\
     $\kappa_{l}$
     & 0.1167 & 0.1169 &  0.1168  &  0.1169
     \\
     $\kappa_{q}$
     & 0.900 & 0.911 &  0.624  &  0.911
     \\
     \hline
     $\chi^{2}$
      & 2.63  & 4.31  &  3.70  &  4.25
     \\
     $\Omega_c h^2$
     & 0.1199  & 0.1199  &  0.1199  &  0.1199
     \\
     $\sigma^{SI}(\times 10^{-44} cm^2)$
     & 0.0967  & 0.1761    &  0.1042   &  0.1596
     \\
     \hline
     $M_{A_H} (GeV)$
     & 169.81 & 117.50 &  158.40  &  117.50
     \\
     $M_{W_H}$ ($M_{Z_H}$)  (GeV)
     & 701.99 & 497.75 &  657.08  &  497.75
     \\
     $M_{\Phi}$  (GeV)
     & 805.19 & 572.15 & 753.91  &  572.15
     \\
     $M_T$  (GeV)
     & 1918.32 & 1137.19 &  1841.03  &  1137.19
     \\
     $M_{T_-}$  (GeV)
     & 899.30 & 747.19 &  1660.50  &  747.19
     \\
     $M_{\nu_{-}}$  (GeV)
     & 183.98 & 130.66 &  172.41  &  130.66
     \\
     $M_{l_{-}}$  (GeV)
     & 185.09 & 132.22 &  173.60  &  132.22
     \\
     $M_{d_{-}}$  (GeV)
     & 1427.05 & 1030.40 &  926.38  &  1030.40
     \\
     $M_{u_{-}}$  (GeV)
     & 1418.45 & 1018.20 &  920.02  &  1018.20
     \\
     \hline
     $|C_{hgg}/SM|^2$
     & 0.861 & 0.734 &  0.842  &  0.734
     \\
     $|C_{hbb}/SM|^2$
     & 0.977 & 0.957 &  0.864  &  0.769
     \\
     $|C_{h\tau\tau}/SM|^2$
     & 0.977 & 0.957 &  0.864  &  0.769
     \\
     $|C_{h\gamma\gamma}/SM|^2$
     & 0.985 & 0.972 & 0.983  &  0.972
     \\
     $|C_{hWW}/SM|^2$
     & 0.976 & 0.953 & 0.973  &  0.953
     \\
     $|C_{hZZ}/SM|^2$
     & 0.976 & 0.953 & 0.973  &  0.953
     \\
     $|C_{htt}/SM|^2$
     & 0.946 & 0.907 & 0.935  &  0.907
     \\
     \hline
     LHC, ggF+ttH, $\gamma\gamma$
     & 0.772 & 0.580 & 0.861  &  0.758
     \\
     LHC, VBF+VH, $\gamma\gamma$
     & 0.991 & 0.978 & 1.147  &  1.277
     \\
     LHC, ggF+ttH, $ZZ^*$
     & 0.757 & 0.558 & 0.842  &  0.728
     \\
     LHC, VBF+VH, $ZZ^*$
     & 0.972 & 0.940 & 1.122  &  1.227
     \\
     LHC, ggF+ttH, $WW^*$
     & 0.757 & 0.558 & 0.842  &  0.728
     \\
     LHC, VBF+VH, $WW^*$
     & 0.972 & 0.940 & 1.122  &  1.227
     \\
      LHC,VH, $b\bar{b}$
      & 0.974 & 0.948 & 0.886 & 0.799
     \\
      LHC, ggF+ttH, $\tau\tau$
      & 0.759 & 0.563 & 0.665  &  0.474
     \\
      LHC, VBF+VH, $\tau\tau$
      & 0.974 & 0.948 & 0.886  &  0.799
     \\
  \hline \hline
  \end{tabular}
\label{best}\vspace{-0.35cm}
\end{table}

\section{Conclusion}
The LHT provides a heavy photon as a candidate for weakly
interacting massive particle dark matter. In this note we examined
the status of such a dark matter candidate in light of the new
results from the LHC Higgs search, the Planck dark matter relic
density and the XENON100 limit on the dark matter scattering off the
nucleon. We scaned over the parameter space in the ranges allowed by
the electroweak precision data. By confronting the parameter space
with the LHC Higgs data, we found that the LHT can well be consistent
with the CMS results but disfavored by the ATLAS observation of
diphoton enhancement. Then in the parameter space allowed by
the CMS Higgs data at $2\sigma$ level, we calculated the  heavy photon
 relic density and found that the heavy photon can
account for the Planck data for the small mass splitting of mirror
lepton and heavy photon.  Finally, under the constraints from LHC
Higgs and Planck dark matter relic density,
 we checked the heavy photon scattering off the nucleon
and found that the heavy photon can give a spin-independent cross section
below the XENON100 upper limit for $m_{A_H}>95$ GeV ($f> 665$ GeV).
The whole parameter space allowed by the current experiments (LHC Higgs,
Planck, XENON100) can be covered by the future XENON1T (2017) experiment
for a $f$ value up to 3.8 TeV.

\section*{Acknowledgment}
Lei Wang would like to thank Genevieve Belanger for helpful discussion
on MicrOMEGAs. This work was supported by the National Natural Science
Foundation of China (NNSFC) under grant Nos. 11005089, 11105116,
11275245, 10821504 and 11135003.

\end{document}